\documentclass{article}
\newcommand{\C}[1]{\mbox{$\{{#1}\}$}}           
\newcommand{\la}{\mbox{$\:\leftarrow\:$}}
\newcommand{\ra}{\mbox{$\:\rightarrow\:$}}
\newcommand{\LL}{\mbox{$\ldots$}}
\newcommand{\NI}{\noindent}
\newcommand{\VV}{\vspace{5 mm}}
\newcommand{\II}{\vspace{2 mm}}
\newcommand{\vect}[1]{{\bf #1}}
\newcommand{\Seq}[1]{{\bf #1}}

\usepackage{latexsym}
\usepackage{alltt}

\begin{document}
\date{}

\title{The Logic Programming Paradigm and Prolog}
\author{Krzysztof R. Apt}

\maketitle

\begin{abstract}
  This is a tutorial on logic programming and Prolog appropriate for a
  course on programming languages for students familiar with
  imperative programming.
\end{abstract}

\tableofcontents

\newpage

\section{History of Logic Programming}
\label{sec:history}

The logic programming paradigm has its roots in automated theorem
proving from which it took the notion of a deduction. What is new is
that in the process of deduction some values are computed.  The
creation of this programming paradigm is the outcome of a long history
that for most of its course ran within logic and only later inside
computer science.  Logic programming is based on the syntax of
first-order logic, that was originally proposed in the second half of
19th century by Gottlob Frege and later modified to the currently used
form by Giuseppe Peano and Bertrand Russell.

In the 1930s Kurt G\"{o}del and Jacques Herbrand studied the notion of
computability based on derivations. These works can be viewed as the
origin of the ``computation as deduction'' paradigm.  Additionally,
Herbrand discussed in his PhD thesis a set of rules for manipulating
algebraic equations on terms that can be viewed now as a sketch of a
unification algorithm.
Some thirty years later in 1965 Alan Robinson published his
fundamental paper \cite{Rob65} that lies at the foundations of the field of
automated deduction.  In this paper he introduced the resolution
principle, the notion of unification and a unification algorithm.
Using the resolution method one can prove theorems of first-order
logic, but another step was needed to see how one could compute within
this framework.  

This was eventually achieved in 1974 by Robert Kowalski
\index{Kowalski, R.A.}  in his paper \cite{Kow74} in which logic
programs with a restricted form of resolution were introduced. The
difference between this form of resolution and the one proposed by
Robinson is that the syntax is more restricted, but proving now has a
side effect in the form of a satisfying substitution. This
substitution can be viewed as a result of a computation and
consequently certain logical formulas can be interpreted as programs.
In parallel, Alain Colmerauer \index{Colmerauer, A.}  with his
colleagues worked on a programming language for natural language
processing based on automated theorem proving.  This ultimately led to
creation of Prolog in 1973.  Kowalski and Colmerauer with his team
often interacted in the period 1971--1973.  This influenced their
views and helped them to crystallize the ideas.

Prolog can be seen as a practical realization of the idea of logic
programs.  It started as a programming language for applications in
natural language processing, but soon after it was found that it can be
used as a general purpose programming language, as well.  A number of
other attempts to realize the computation as deduction paradigm were
proposed around the same time, notably by Cordell Green and Carl
Hewitt, but the logic programming proposal, probably because it was the
simplest and most versatile, became most successful.

Originally, Prolog was implemented by Philippe Roussel,
\index{Roussel, Ph.}  a colleague of Colmerauer, in the form of an
interpreter written in Algol-W.  An important step forward was
achieved by David H. Warren \index{Warren. D.H}
who proposed in 1983 an abstract machine,
now called WAM (Warren Abstract Machine), that consists of a machine
architecture with an instruction set which serves as a target for
machine independent Prolog compilers.  WAM became a standard basis for
implementing Prolog and other logic programming languages.

The logic programming paradigm influenced a number of developments in
computer science. Already in the seventies it led to the creation of
deductive databases that extend the relational databases by providing
deduction capabilities.  A further impetus to the subject came
unexpectedly from the Japanese Fifth Generation Project for
intelligent computing systems (1982--1991) in which logic programming
was chosen as its basis.
More recently, this paradigm led to constraint logic
programming that realizes a general approach to computing in which the
programming process is limited to a generation of constraints
(requirements) and a solution of them, and to inductive logic
programming, a logic based approach to machine learning.

The above account of history of logic programming and Prolog shows its
roots in logic and automated deduction. In fact, Colmerauer and
Roussel write in \cite{CR96}: ``There is no question that Prolog is
essentially a theorem prover `\`{a} la Robinson.'  Our contribution
was to transform that theorem prover into a programming language.''
This origin of the logic paradigm probably impeded its acceptance
within computer science in times when imperative programming 
got impetus thanks to the creation of Pascal and C, the fields of
verification and semantics of imperative programs gained ground and
when the artificial intelligence community already adopted Lisp as
the language of their choice.

Here we offer an alternative presentation of the subject by focusing
on the ordinary programming concepts (often implicitly) present in
logic programming and by relating various of its ideas to those
present in the imperative and functional programming paradigms.

\section{Brief Overview of the Logic Programming Paradigm}

The logic programming paradigm substantially differs from other
programming paradigms.  When stripped to the bare essentials it can be
summarized by the following three features:

\begin{itemize}

\item computing takes place over the domain of all terms 
defined over a ``universal'' alphabet.

\item values are assigned to variables by means of automatically generated
substitutions, called {\em most general unifiers}. These values may
contain variables, called \emph{logical variables},

\item the control is provided by a single mechanism: automatic {\em backtracking}.

\end{itemize}

In our exposition of this programming paradigm we shall stress the
above three points.  Even such a brief summary shows both the strength
and weakness of the logic programming paradigm.  Its strength lies in
an enormous simplicity and conciseness; its weakness has to do with
the restrictions to one control mechanism and the use of a single data
type.

So this framework has to be modified and enriched to accommodate it
to the customary needs of programming, for example by
providing various control constructs and by introducing the data type of
integers with the customary arithmetic operations.  This can be done
and in fact Prolog and constraint logic programming languages are
examples of such a customization of this framework.

\paragraph{Declarative programming} 
Two additional features of logic programming are important to note.
First, in its pure form it supports {\em declarative programming}.  A
{\em declarative program\/} admits two interpretations.  The first
one, called a {\em procedural interpretation\/}, \index{procedural
  interpretation} explains {\em how\/} the computation takes place,
whereas the second one, called a {\em declarative interpretation\/},
\index{declarative interpretation} is concerned with the question {\em
  what\/} is being computed.

Informally, the procedural interpretation is concerned with the {\em
  method}, whereas the declarative interpretation is concerned with
the {\em meaning}.  In the procedural interpretation a declarative
program is viewed as a description of an algorithm that can be
executed.  In the declarative interpretation a declarative program is
viewed as a formula, and one can reason about its correctness without
any reference to the underlying computational mechanism.  This makes
declarative programs easier to understand and to develop.

As we shall see, in some situations the specification of a problem in
the logic programming format already forms an algorithmic solution to
the problem. So logic programming supports declarative programming and
allows us to write {\em executable specifications}. It should be added
however, that in practice the Prolog programs obtained in this way are
often inefficient, so this approach to programming has to be combined
with various optimization techniques, and an appropriate understanding
of the underlying computation mechanism is indispensable.  To clarify
this point we shall present here a number of Prolog programs that are
declarative and eliminate from them various sources of inefficiency.

This dual interpretation of declarative programs also accounts for the
double use of logic programming ---as a formalism for programming and
for knowledge representation, and explains the importance of logic
programming in the field of artificial intelligence.

\paragraph{Interactive Programming}
Another important feature of logic programming is that it supports
{\em interactive programming}.  That is, the user can write a single
program and interact with it by means of various queries of interest
to which answers are produced.  The Prolog systems greatly support
such an interaction and provide simple means to compute one or more
solutions to the submitted query, to submit another query, and to
trace the execution by setting up, if desired, various check points,
all within the same ``interaction loop''.  This leads to a flexible
style of programming.

This is completely analogous to the way functional programs are used
where the interaction is achieved by means of expressions that need to
be evaluated using a given collection of function definitions.

\medskip

In what follows we shall introduce Prolog, the best known programming
language based on the logic programming paradigm. Prolog is then based
on a subset of first-order logic. We explain here how Prolog uses this
syntax in a novel way (this characteristic is called \emph{ambivalent
  syntax}) and extends it by a number of interesting features, notably
by supporting infix notation and by providing so-called
\emph{anonymous} and \emph{meta-variables}. These extensions amount to
more than syntactic sugar.  In fact, they make it possible to realize
in Prolog higher-order programming and meta-programming in a simple
way.

When discussing Prolog it is useful to abstract from the programming
language and first consider the underlying conceptual model provided
by logic programming.

\section{Equations Solved by Unification as Atomic Actions}
\label{sec:unification}

We begin by explaining how computing takes place at the ``atomic
level''.  In logic programming the atomic actions are equations
between terms (arbitrary expressions).  They are executed by means of
the unification process that attempts to solve such equations. In the
process of solving values are assigned to variables. These values can
be arbitrary terms.  In fact, the variables are all of one type that
consists of the set of all terms.

This informal summary shows that the computation process in 
logic programming is governed by different principles
than in the other programming paradigms.

\subsection{Terms}
\label{subsec:terms}

In a more rigorous explanation let us start by introducing an alphabet
that consists of the following disjoint classes of symbols:
\begin{itemize}
\item
{\em variables\/}, denoted by $x,y,z,\dots$ possibly with subscripts,
\item
{\em function symbols\/}, \index{function symbol}
\item
{\em parentheses\/}, ``('' and ``)'',
\item
{\em comma\/}, ``,''.
\end{itemize}

We also postulate that each function symbol has a fixed {\em
arity\/}, that is the number of arguments associated
with it.  0-ary function symbols are called
{\em constants\/},
and are usually denoted by $a,b,c,d,\dots$.
Below we denote function symbols of positive arity 
by $f,g,h,\dots$.

Finally, {\em terms\/} are defined inductively as follows:
\begin{itemize}
\item
a variable is a term,
\item
if $f$ is an $n$-ary function symbol and $t_1,\dots,t_n$ are terms, then
$f(t_1,\dots,t_n)$ is a term.
\end{itemize}
In particular every constant is a term.  Variable-free terms are usually
called {\em ground terms}. 
Below we denote terms by
$s,t,u,w, \dots$.  

For example, if $a$ is a constant, $x$ and $y$ are variables,
$f$ is a binary function symbol and $g$ a unary function symbol, then
$f(f(x,g(b)),y)$ is a term.

Terms are fundamental concepts in mathematical logic but at first
sight they seem to be less common in computer science.  However, they
can be seen as a generalization of the concept of a string familiar
from the theory of formal languages. In fact, strings can be viewed as
terms built out of an alphabet the only function symbols of which are
the concatenation operations in each arity (or alternatively, out of
an alphabet the only function symbol of which is the binary
concatenation operation assumed to be associative, say to the right).
Another familiar example of terms are arithmetic expressions. These
are terms built out of an alphabet in which as the function symbols we
take the usual arithmetic operations of addition, subtraction,
multiplication, and, say, integer division, and as constants 0, -1, 1, \LL .

In logic programming no specific alphabet is assumed. In fact, it is
convenient to assume that in each arity an infinite supply of function
symbols exists and that all terms are written in this ``universal
alphabet''.  These function symbols can be in particular the
denotations of arithmetic operations but no meaning is attached to
these function symbols. This is in contrast to most of the imperative
programming languages, in which for example the use of ``+'' in an
expression implies that we refer to the addition operation.  The other
consequence of this choice is that no types are assigned to terms.  In
fact, no types are assumed and consequently there is no distinction
between, say, arithmetic expressions, Boolean expressions, and terms
denoting lists.  All these terms are considered as being of one type.

\subsection{Substitutions}
\label{subsec:substitutions}

Unlike in imperative programming, in logic programming the variables
can be uninitialized.  Moreover, the possible values of variables are
terms. So to explain properly the computation process we need to
reexamine the notion of a state.

At any moment during the computation there will be only a finite
number of variables that are \emph{initialized} ---these are
variables to which in the considered computation some value was
already assigned. Since these values are terms, we are naturally led
to consider \emph{substitutions}. These are finite mappings from
variables to terms such that no variable is mapped to itself.  So
substitution provides information about which variables are initialized.
(Note that no variable can be initialized to itself which explains the
restriction that no variable is mapped to itself.)

Substitutions then form a counterpart of the familiar notion of a
{\em state\/} used in imperative programming.
We denote a substitution by $\C{x_1 / t_1, \dots, x_n / t_n}$.
This notation implies that $x_1, \dots, x_n$ are different variables,
$t_1, \dots, t_n$ are terms and that no term $t_i$ equals the variable $x_i$.
We say then that the substitution $\C{x_1 / t_1, \dots, x_n / t_n}$
\emph{binds} the variable $x_i$ to the term $t_i$. 

Using a substitution we can evaluate a term in much the same way as
using a state we can evaluate an expression in imperative programming
languages.  This process of evaluation is called an \emph{application}
of a substitution to a term.  It is the outcome of a simultaneous
replacement of each variable occurring in the domain of the
substitution by the corresponding term.  So for example the
application of the substitution $\C{x/f(z), y/g(z)}$ to the term
$h(x,y)$ yields the term $h(f(z),g(z))$. Here the variable $x$ was
replaced by the term $f(z)$ and the variable $y$ by the term $g(z)$.
In the same way we define an application of a substitution to an atom,
query, or a clause.

So an evaluation of a term using a substitution yields again a term.
This is in contrast to imperative programming where an evaluation
of an expression using a state yields a value that belongs to the type
of this expression.  

\subsection{Most General Unifiers}
\label{subsec:mgu}

As already mentioned, in logic programming the atomic actions are equations
between terms and the unification process is used to determine their meaning.
Before we discuss these matters in detail let us consider some
obvious examples of how solving equations can be used as an assignment.

We assume that all mentioned variables are uninitialized.  By writing
$x = a$ we assign the constant $a$ to the variable $x$. Since in logic
programming the equality ``='' is symmetric, the same effect is
achieved by writing $a = x$.  More interestingly, by writing $x =
f(y)$ (or, equivalently, $f(y) = x$) we assign the term $f(y)$ to the
variable $x$. Since $f(y)$ is a term with a variable, we assigned to
the variable $x$ an expression with a variable in it. Recall that a
variable that occurs in a value assigned to another variable is called
a logical variable. So $y$ is a logical variable here.  The use of
logical variables is an important distinguishing feature of logic
programming and we devote the whole Subsection \ref{subsec:logical} to
an explanation of their use.  Finally, by writing $f(y) = f(g(a))$ we
assign the term $g(a)$ to the variable $y$, as this is the way
to make these two terms equal.

These examples show that the equality ``='' in logic programming
and the assignment in C, also written using ``='', are totally
different concepts.

Intuitively, \emph{unification} is the process of solving an equation
between terms (i.e., of making two terms equal) in a least
constraining way. The resulting substitution (if it exists) is called
a \emph{most general unifier} (\emph{mgu}).  For example, the equation $x =
f(y)$ can be solved (i.e., the terms $x$ and $f(y)$ \emph{unify}) in a
number of different ways, for instance by means of each of the
substitutions $\C{x/f(y)}$, $\C{x/f(a), y/a}$, $\C{x/f(a), y/a,
  z/g(b)}$, Clearly only the first one is ``least constraining''. In
fact, out of these three substitutions the first one is the only most
general unifier of the equation $x = f(y)$.
The notion of a least constraining substitution can be made precise by
defining an order on substitutions. In this order the substitution
$\C{x/f(y)}$ is more general than $\C{x/f(a), y/a}$, etc.

Note that we made the terms $x$ and $f(y)$ equal by instantiating only
one of them.  Such a special case of the unification is called
\emph{matching} which is the way of assigning values in functional
programming languages.  Unification is more general than matching as
the following, slightly less obvious, example shows.  Consider the
equation $f(x,a) = f(b,y)$. Here the most general unifier is $\C{x/b,
  y/a}$.  In contrast to the previous example it is now not possible to
make these two terms equal by instantiating only one of them.

The problem of deciding whether an equation between terms has a
solution is called the {\em unification problem}. \index{unification
  problem} Robinson showed in \cite{Rob65} that the unification
problem is decidable. More precisely, he introduced a unification
algorithm with the following property.  If an equation between terms
has a solution, the algorithm produces an mgu and otherwise it reports
a failure.  An mgu of an equation is unique up to renaming of the
variables.

\subsection{A Unification Algorithm} \label{subsec:mm}

In what follows we discuss the unification process in more detail using
an elegant unification algorithm introduced in Martelli
and Montanari \cite{MM82}.
This algorithm takes 
as input a finite set of term equations $\C{s_1 = t_1, \LL, s_n = t_n}$
and tries to produces an mgu of them.
\VV

\NI
{\sc Martelli--Montanari Algorithm}
\II

\NI
Nondeterministically choose from the set of equations an equation of a
form below and perform the associated action.
\begin{tabbing}
\= (2') \=  $x=t$ where $x$ occurs in $t$ and $x$ differs from $t$x \= \kill \\
\> (1) \> $f(s_1 ,...,s_n) = f(t_1 ,...,t_n)$                           \> {\em
replace by the equations}  \\
\>     \>                                                               \> $s_1
= t_1 ,...,s_n = t_n$, \\[2mm]
\> (2) \> $f(s_1 ,...,s_n) = g(t_1 ,...,t_m)$ where $f \neq g$          \> {\em
halt with failure}, \\[2mm]
\> (3) \> $x=x$                                                         \> {\em
delete the equation}, \\[2mm]
\> (4) \> $t=x$ where $t$ is not a variable                             \> {\em
replace by the equation} $x=t$, \\[2mm]
\> (5) \> $x=t$ where $x$ does not occur in $t$                         \> {\em
 apply the substitution \C{x/t}} \\
\>     \> and $x$ occurs elsewhere                                      \> {\em
 to all other equations}\\[2mm]
\> (6) \> $x=t$ where $x$ occurs in $t$ and $x$ differs from $t$        \> {\em
 halt with failure.} 
\end{tabbing}
\medskip

The algorithm terminates when no action can be performed or when
failure arises. In case of success, by changing in the final set of
equations all occurrences of ``$=$'' to ``$/$'' we obtain the desired
mgu.  Note that action (1) includes the case $c=c$ for every constant
$c$ which leads to deletion of such an equation. In addition, action
(2) includes the case of two different constants.  Finally, note that
no action is performed when the selected equation is of the form $x=t$
where $x$ does not occur occur elsewhere (so a fortiori does not occur
in $t$). In fact, in case of success all equations will be of such a
form.

To illustrate the operation of this algorithm reconsider the equation
$f(x,a) = f(b,y)$.  Using action (1) it rewrites to the set of two
equations, $\C{x = b, a = y}$. By action (4) we now get the set 
$\C{x = b, y = a}$. At this moment the algorithm terminates and we obtain the mgu
$\C{x/b, y/a}$.

So by interpreting the equality symbol as the request to find
a most general unifier of the considered pair of terms,
each equation is turned into
an atomic action that either produces a substitution
(a most general unifier) or \emph{fails}.
This possibility of a failure at the level of an atomic
action is another distinguishing feature of logic programming.

By writing a sequence of equations we can create very
simple logic programs that either succeed and produce
as output a substitution or fail.
It is important to understand how the computation then proceeds.
We illustrate it by means of three progressively more complex
examples.

First, consider the sequence 
\[
f(x,a) = f(g(z),y), \ h(u) = h(d).
\]
The first equation yields first the intermediate substitution
$\C{x/g(z), y/a}$ and the second one the substitution $\C{u/d}$. By
combining these two substitutions we obtain the substitution
$\C{x/g(z), y/a, u/d}$ produced by this logic program.

As a slightly less obvious example consider the sequence 
\[
f(x,a) = f(g(z),y), \ h(x,z) = h(u,d).
\]
Here the intermediate substitution $\C{x/g(z), y/a}$ binds the
variable $x$ that also occurs in the second equation.  This second
equation needs to be evaluated first in the ``current state'', here
represented by the substitution $\C{x/g(z), y/a}$, before being
executed. This evaluation produces the equation $h(g(z),z) = h(u,d)$.
This equation yields the most general unifier $\C{u/g(d), z/d}$ and
the resulting final substitution is here $\C{x/g(d), y/a, u/g(d),
  z/d}$.

What happened here is that the substitution $\C{u/g(d), z/d}$ was
\emph{applied} to the intermediate substitution $\C{x/g(z), y/a}$.
The effect of an application of one substitution, say
$\delta$, to another, say $\gamma$,
(or of \emph{composition} of the substitutions) is obtained by 
\begin{itemize}
\item applying 
$\delta$ to each of the terms that appear in the range of $\gamma$,

\item adding to the resulting substitution the bindings to the 
variables that are in the domain of $\delta$ but not in the domain of $\gamma$.
\end{itemize}
In the above example the first step yields the substitution
$\C{x/g(d), y/a}$ while the second step adds the bindings $u/g(d)$ and
$z/d$ to
the final substitution.  This process of substitution composition
corresponds to an \emph{update of a state} in imperative programming
and that is how we shall refer to it in the sequel.

As a final example consider the sequence  
\[
f(x,a) = f(g(z),y), \ h(x,z) = h(d,u).
\]
It yields a failure. Indeed, after executing the first equation the
variable $x$ is bound to $g(z)$, so the evaluation of the second
equation yields $h(g(z),z) = h(d,u)$ and no substitution makes equal
(unifies) the terms $h(g(z),z)$ and $h(d,u)$.

It is useful to compare solving equations by unification with the
assignment command.  First, note that, in contrast to assignment,
unification can assign an arbitrary term to a variable. Also it
can fail, something the assignment cannot do.  On the other hand, using
assignment one can modify the value of a variable, something
unification can perform in a very limited way: by further
instantiating the term used as a value.  So these atomic actions are
incomparable.

\section{Clauses as Parts of Procedure Declarations}
\label{sec:lp}

Logic programming is a rule based formalism and Prolog is a rule based language.
In this context the rules are called clauses.
To better understand the relationship between logic programming and
imperative programming we proceed in two steps and introduce
a restricted form of clauses first.

\subsection{Simple Clauses}
\label{subsec:simple-clauses}

Using unification we can execute only extremely simplistic programs
that consist of sequences of equations.  We now enrich this framework
by adding procedures.  In logic programming they are modelled by means
of {\em relation symbols}, \index{relation symbol} sometimes called
{\em predicates\/}.  Below we denote relation symbols by
$p,q,r,\dots$.  As in the case of the function symbols, we assume that
each relation symbol has a fixed arity associated with it. When the
arity is 0, the relation symbol is usually called a {\em propositional
  symbol}. \index{propositional symbol}

If $p$ is an $n$-ary relation symbol and $t_1,\dots,t_n$ are terms,
then we call $p(t_1,\dots,t_n )$ an {\em atom}. \index{atom} When $n =
0$ the propositional symbols coincide with atoms.  Interestingly, as
we shall see, such atoms are useful.  Intuitively, a relation symbol
corresponds to a \emph{procedure identifier} and an atom to a
\emph{procedure call}.  The equality symbol ``='' is a binary relation
symbol written in an infix form, so each equation is also an atom.
However, the meaning of equality is determined, so it can be viewed as
a built-in procedure, i.e., a procedure with a predefined meaning.

We still need to define the procedure declarations
and to clarify the parameter mechanism used.
Given an $n$-ary relation symbol $p$ and atoms $A_1, \LL, A_k$ we call
an expression of the form
\[
p(x_1, \LL, x_n) :- \: A_1, \LL, A_k.
\]
a \emph{simple clause}. \index{simple clause} $p(x_1, \LL, x_n)$ is called
the \emph{head} of the clause and $A_1, \LL, A_k$ its \emph{body}.
The fullstop ``.'' at the end of the clause is important: it
signals to the compiler (or interpreter) that the end of the clause is
encountered.

The procedural interpretation of a simple clause $p(x_1, \LL, x_n) :-
\: A_1, \LL, A_k$ is: ``to establish $p(x_1, \LL, x_n)$ establish
$A_1, \LL, A_k$'', while the declarative interpretation is: ``$p(x_1,
\LL, x_n)$ is true if $A_1, \LL, A_k$ is true''.  The declarative
interpretation explains why in the logic programming theory the reversed
implication symbol ``$\la$'' is used instead of ``{\tt :-}''.

Finally a \emph{simple logic program} \index{simple logic program} is
a finite set of clauses.  Such a program is activated by providing an
initial \emph{query}, \index{query} which is a sequence of atoms.  In
the imperative programming jargon a query is then a program and a
simple logic program is a set of procedure declarations.  Intuitively, given
a simple program, the set of its simple clauses with the same relation symbol
in the head corresponds to the procedure declaration in the imperative
languages.  One of the syntactic confusions is
that in logic programming the comma ``,'' is used as a separator
between the atoms constituting a query, whereas in the imperative
programming the semicolon ``;'' is used for this purpose.

\subsection{Computation Process}
\label{subsec:computing}

A nondeterminism is introduced into this framework by
allowing \emph{multiple clauses} with the same relation symbol in the head.
In the logic programming theory this form of nondeterminism (called
{\em don't know nondeterminism\/}) \index{don't know nondeterminism}
is retained by considering all computations that can be generated by 
means of multiple clauses and by retaining the ones that lead to
a success. ``Don't know'' refers to the fact that in general we do not know
which computation will lead to a success.

In Prolog this computation process is made deterministic by ordering
the clauses by the textual ordering and by employing automatic
backtracking to recover from failures.  Still, when designing Prolog
programs it is useful to have the don't know nondeterminism in mind.
In fact, in explanations of Prolog programs phrases like ``this
program nondeterministically guesses an element such that \LL'' are
common.  Let us explain now more precisely how the computing takes
place in Prolog.  To this end we need to clarify the procedure
mechanism used and the role played by the use of multiple clauses.

The procedure mechanism associated with the simple clauses introduced
above is \emph{call-by-name} according to which the formal parameters
are simultaneously substituted by the actual ones. So this procedure
mechanism can be simply explained by means of substitutions: given a
simple clause $p(x_1, \LL, x_n) :- \: A_1, \LL, A_k .$ a procedure
call $p(t_1,\dots,t_n )$ leads to an execution of the statement $(A_1,
\LL, A_k) \C{x_1 / t_1, \dots, x_n / t_n}$ obtained by applying the
substitution $\C{x_1 / t_1, \dots, x_n / t_n}$ to the statement $A_1,
\LL, A_k$.
(We assume here that the
variables of the clauses are appropriately renamed to avoid variable
clashes.)  Equivalently, we can say that the procedure call
$p(t_1,\dots,t_n)$ leads to an execution of the statement $A_1, \LL,
A_k$ in the state (represented by a substitution) updated by 
the substitution $\C{x_1 / t_1, \dots, x_n / t_n}$.

The clauses are tried in the order they appear in the program text.
The depth-first strategy is implied by the fact that a procedure call
leads directly to an execution of the body of the selected simple
clause.  If at a certain stage a failure arises, the computation
backtracks to the last choice point (a point in the computation at
which one out of more applicable clauses was selected) and the
subsequent simple clause is selected.  If the selected clause was the
last one, the computation backtracks to the previous choice point. If
no choice point is left, a failure arises.  Backtracking implies that
the state is restored, so all the state updates performed since
the creation of the last choice point are undone.

Let us illustrate now this definition of Prolog's computation
process by considering the most known Prolog program the purpose of
which is to append two lists.  In Prolog the empty list is denoted by
\texttt{[]} and the list with head \texttt{h} and tail \texttt{t} by
\texttt{[h | t]}. The term \texttt{[a | [b | s]]} abbreviates to a
more readable form \texttt{[a,b | s]}, the list \texttt{[a | [b |
  []]]} abbreviates to \texttt{[a,b]} and similarly with longer
lists. This notation can be used both for lists and for arbitrary
terms that start with the list formation operator \texttt{[.|..]}.

Then the following logic program defines by induction w.r.t. the first
argument how to append two lists. Here and elsewhere we follow
Prolog's syntactic conventions and denote variables by strings
starting with an upper case letter.  The names ending with ``{\tt s}''
are used for the variables meant to be instantiated to lists.

\begin{verbatim}
% append(Xs, Ys, Zs) :- Zs is the result of concatenating 
%                       the lists Xs and Ys. 
  append(Xs, Ys, Zs) :- Xs = [], Zs = Ys.
  append(Xs, Ys, Zs) :- Xs = [H | Ts], Zs = [H | Us], append(Ts, Ys, Us).
\end{verbatim}

In Prolog the answers are generated as substitutions written in an equational
form (as in the Martelli--Montanari algorithm presented above).
In what follows we display a query $Q.$ as \texttt{?-} $Q$. . Here
``?-'' is the system prompt and the fullstop ``.'' 
signals the end of the query.

One can check then that the query 
\medskip

\noindent
\texttt{?- append([jan,feb,mar], [april,may], Zs).}
\medskip

\noindent
yields \texttt{Zs = [jan,feb,mar,april,may]} as the answer and that
the query 
\medskip

\noindent
\texttt{?- append([jan,feb,mar], [april,may], [jan,feb,mar,april,may]).}
\medskip

\noindent
succeeds and yields the empty substitution as the answer.
In contrast, the query 
\medskip

\noindent
\texttt{?- append([jan,feb,mar], [april,may], [jan,feb,mar,april]).} 
\medskip

\noindent
fails. Indeed, the computation leads to the subsequent procedure calls
\medskip

\texttt{append([feb,mar], [april,may], [feb,mar,april])},

\texttt{append([mar], [april,may], [mar,april])} and

\texttt{append([], [april,may], [april])}, 
\medskip

\noindent
and the last one fails
since the terms \texttt{[april,may]} and \texttt{[april]} don't
unify. 

\subsection{Clauses}
\label{subsec:clauses}

The last step in defining logic programs consists of allowing
arbitrary atoms as heads of the clauses. 
Formally, given atoms $H, A_1, \LL, A_k$, we call
an expression of the form
\[
H :- \: A_1, \LL, A_k.
\]
a \emph{clause}. \index{clause} If $k=0$, that is if the clause's body
is empty, such a clause is called a \emph{fact} \index{fact} and the
``{\tt :-}'' symbol is then omitted.  If $k>0$, that is, if the
clause's body is non-empty, such a clause is called a \emph{rule}.
\index{rule} A \emph{logic program} \index{logic program} is then a
finite set of clauses and a \emph{pure Prolog program} is a finite
sequence of clauses.

Given a pure Prolog program, we call the set of its clauses with
the relation $p$ in the head the \emph{definition of $p$}.
Definitions correspond to the procedure declarations in imperative
programming and to the function definitions in functional
programming.
Variables that occur in the body of a clause but not in its head are
called \emph{local}. They correspond closely to the variables that are
local to the procedure bodies in the imperative languages with the
difference that in logic programs their declaration is implicit.
Logic programming, like Pascal, does not have a block statement.

To explain how the computation process takes place for pure Prolog
programs we simply view a clause of the form
\[
p(s_1, \LL, s_n) :- \: A_1, \LL, A_k.
\]
as a shorthand for the simple clause
\[
p(x_1, \LL, x_n) :- \: (x_1, \LL, x_n) = (s_1, \LL, s_n), A_1, \LL, A_k.
\]
where $x_1, \LL, x_n$ are fresh variables.  We use here Prolog's
syntactic facility according to which given a sequence $s_1,\dots,s_n$
of terms $(s_1,\dots,s_n )$ is also a term.

So given a procedure call $p(t_1,\dots,t_n )$ if the above clause 
$p(s_1, \LL, s_n) :- \: A_1, \LL, A_k$ is
selected, an attempt is made to unify $(t_1,\dots,t_n )$ with $(s_1,
\LL, s_n)$.  (As before we assume here that no variable clashes arise.
Otherwise the variables of the clause should be appropriately renamed.)
If the unification succeeds and produces a substitution $\theta$,
the state (represented by a substitution) is updated by applying to it
$\theta$ and the computation continues with the statement $A_1, \LL,
A_k$ in this new state. Otherwise a failure arises and the next clause
is selected.

So due to the use of clauses instead of simple clauses, unification is
effectively lifted to a parameter mechanism.  As a side effect this
makes the explicit use of unification, modelled by means of ``='',
superfluous.  As an example reconsider the above program appending two
lists.  Using the clauses it can be written in a much more
succinct way, as the following program \texttt{APPEND}:

\begin{verbatim}
% append(Xs, Ys, Zs) :- Zs is the result of concatenating 
%                       the lists Xs and Ys. 
  append([], Ys, Ys).
  append([X | Xs], Ys, [X | Zs]) :- append(Xs, Ys, Zs).
\end{verbatim}
Here the implicit case analysis present in the previous program is in
effect moved into the heads of the clauses. 
The use of terms in the heads of the clauses is completely
analogous to the use of \emph{patterns} in function definitions in functional
programming.

To summarize, the characteristic elements of procedure declarations in
logic programming, in contrast to imperative programming, are: the use
of multiple rules and use of patterns to select among these rules.

\section{Prolog's Approach to Programming}

The power and originality of the Prolog programming style lies in the combination of
automatic backtracking with the use of relations and 
logical variables.

\subsection{Multiple Uses of a Single Program}
\label{subsec:multiple}

As a first illustration of the novelty of Prolog's approach to programming
we illustrate the possibility of using the same program for different purposes.
The perhaps simplest example involves the following program {\tt MEMBER}.
We use in it a useful feature of Prolog, so-called {\em anonymous
  variable\/}, \index{variable!anonymous} written as an
``underscore'' character ``\_'' .  Each occurrence of ``\_'' in a
query or in a clause is interpreted as a {\em different\/} variable.
Anonymous variables are analogous to the \emph{wildcard pattern}
feature of the Haskell language.  

\begin{verbatim}
% member(X, Xs):-  X is a member of the list Xs.
  member(X, [X | _]).
  member(X, [_ | Xs]):-  member(X, Xs).
\end{verbatim}

{\tt MEMBER} can be used both for testing and for computing:

\begin{verbatim}
?- member(wed, [mon, wed, fri]).

yes

?-  member(X, [mon, wed, fri]).

Xs = mon ;

Xs = wed ;

Xs = fri ;

no
\end{verbatim}
Here ``;'' is the user's request to produce the next answer. If this
request fails, the answer ``{\tt no}'' is printed.

Consequently, given a variable {\tt X} and two lists {\tt s} and {\tt t},
the query {\tt member(X, s), member(X, t).} generates all elements
that are present both in {\tt s} and {\tt t}.
Operationally, the first call generates all members of {\tt s}
and the second call tests for each of them the membership in {\tt t}.

Also the \texttt{APPEND} program can be used for a number of purposes,
in particular to concatenate two lists and to split a list in all
possible ways. For example we have

\begin{verbatim}
?- append(Xs, Ys, [mon, wed, fri]).

Xs = []
Ys = [mon, wed, fri] ;

Xs = [mon]
Ys = [wed, fri] ;

Xs = [mon, wed]
Ys = [fri] ;

Xs = [mon, wed, fri]
Ys = [] ;

no
\end{verbatim}

This cannot be achieved with any functional programming version of the
{\tt APPEND} procedure.  The difference comes from the fact that in
logic programming procedures are defined by means of relations
whereas in functional programming functions are used. In fact, there
is no distinction between input and output arguments in the procedures
in logic programs.

To see two uses of {\tt append} in a single program consider a program
that checks whether one list is a consecutive sublist of another one.
The one line program \texttt{SUBLIST} that follows formalizes the following
definition of a sublist:
\begin{itemize}
\item the list {\tt Xs} is a sublist of the list {\tt Ys} if {\tt Xs}
is a prefix of a suffix of {\tt Ys}.
\end{itemize}

\begin{verbatim}
% sublist(Xs, Ys) :- Xs is a sublist of the list Ys.
  sublist(Xs, Ys) :- append(_, Zs, Ys), append(Xs, _, Zs).
\end{verbatim}

\begin{figure}[htbp]
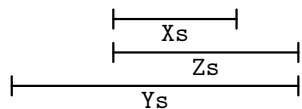

\medskip

\catcode`@=11
\expandafter\ifx\csname graph\endcsname\relax \alloc@4\box\chardef\insc@unt\graph\fi
\expandafter\ifx\csname graphtemp\endcsname\relax \alloc@1\dimen\dimendef\insc@unt\graphtemp\fi
\catcode`@=12
\setbox\graph=\vtop{%
  \vbox to0pt{\hbox{%
    \special{pn 8}%
    \special{pn 13}%
    \special{pa 1500 400}%
    \special{pa 0 400}%
    \special{fp}%
    \special{pa 0 450}%
    \special{pa 0 350}%
    \special{fp}%
    \special{pa 1500 450}%
    \special{pa 1500 350}%
    \special{fp}%
    \graphtemp=.6ex \advance\graphtemp by 0.48in
    \rlap{\kern 0.75in\lower\graphtemp\hbox to 0pt{\hss \tt Ys\hss}}%
    \special{pa 1500 225}%
    \special{pa 533 225}%
    \special{fp}%
    \special{pa 533 275}%
    \special{pa 533 175}%
    \special{fp}%
    \special{pa 1500 275}%
    \special{pa 1500 175}%
    \special{fp}%
    \graphtemp=.6ex \advance\graphtemp by 0.305in
    \rlap{\kern 1.016in\lower\graphtemp\hbox to 0pt{\hss \tt Zs\hss}}%
    \special{pa 533 50}%
    \special{pa 1178 50}%
    \special{fp}%
    \special{pa 533 100}%
    \special{pa 533 0}%
    \special{fp}%
    \special{pa 1178 100}%
    \special{pa 1178 0}%
    \special{fp}%
    \graphtemp=.6ex \advance\graphtemp by 0.13in
    \rlap{\kern 0.855in\lower\graphtemp\hbox to 0pt{\hss \tt Xs\hss}}%
    \kern 1.5in
  }\vss}%
  \kern 0.616in
}
\centerline{\box\graph}
\caption{Xs is a sublist of the list Ys \label{fig:sublist} }
\end{figure}

Here both anonymous variables and {\tt Zs} are local.
In this rule {\tt Zs} is a suffix of {\tt Ys} and {\tt Xs} is a prefix
of {\tt Zs}.  This relation is illustrated in Figure
\ref{fig:sublist}. 

Operationally, given two lists, \texttt{as} and
\texttt{bs}, the query \texttt{sublist(as, bs).} leads to a generation
of splits of the list \texttt{bs} through the call 
\texttt{append(\_,  Zs, bs)}.  Then for each generated suffix \texttt{Zs} 
of \texttt{bs} it is checked whether for some list, denoted by the anonymous 
variable \_ , the call \texttt{append(as, \_, Zs)} succeeds.
This happens when \texttt{as} is a prefix of \texttt{Zs}.
So a typical use of this program involves backtracking.

\subsection{Logical Variables}
\label{subsec:logical}

Let us return now to the logical variables. They are an important
feature of logic programming but it is easy to overlook their use. For
example, they already appear in the computations involving the first
version of the list concatenation program, and consequently, because
of the way we defined the computation process, in the computations of
the {\tt APPEND} program. Indeed, given the query {\tt
  append([jan,feb,mar], [april,may], Zs).} the rule 
\begin{verbatim}
  append(Xs, Ys, Zs) :- Xs = [H | Ts], Zs = [H | Us], append(Ts, Ys, Us).  
\end{verbatim}
leads to the binding of the variable {\tt Zs} to the term {\tt [jan |
  Us]}.  The value of the variable {\tt Us} is computed later, by
means of the call {\tt append([feb,mar], [april,may], Us)}. This call
first binds {\tt Us} to the term {\tt [feb | U1s]}, where {\tt U1s} is
a fresh variable, and hence {\tt Zs} to the term {\tt [jan, feb |
  U1s]}.  This progressive building of the output using the logical
variables is typical for Prolog.

The real power of logical variables should become apparent after
considering the following three original Prolog programs.

\subsubsection*{A Type Assignment}

Consider the typed lambda calculus and Curry's system of type
assignment. It involves statements of the form $s: \tau$ which should be
read as ``term $s$ has type $\tau$''.  Finite sequences of such
statements with $s$ being a variable are called {\em environments\/}
are denoted below by $E$.  A statement of the form $E \vdash s: \tau$
should be read as ``in the environment $E$ the term $s$ has type $\tau$''.
The following three rules define by induction on the structure of
lambda terms how to assign types to lambda terms: 
\medskip

\[ \frac{ x:t \in E                }
        { E \vdash  x:t            }
\]

\[ \frac{ E \vdash m : s\ra t \ , \ \  E \vdash n : s}
        { E \vdash  (m \ n) : t}\]

\[ \frac{ E \; , \; x:s \vdash  m:t        }
        { E \vdash  (\lambda x. m) : s\ra t}\]

\VV

To encode the lambda terms as usual ``first-order'' terms we use
the unary function symbol {\tt var} and two binary function symbols,
{\tt lambda} and {\tt apply}.  The lambda term $x$ (a variable) is
translated to the term {\tt var(x)}, the lambda term $(m \ n)$ to the
term {\tt apply(m, n)}, and the lambda term $\lambda x. m$ to the term
{\tt lambda(x, m)}.  For example, the lambda term $\lambda x. \: (x \ 
x)$ translates to {\tt lambda(x, apply(var(x), var(x)))}.  The subtle
point is that according to Prolog convention, lower case letters stand
for constants, so for example {\tt var(x)} is a ground term (i.e. a
term without variables).

The above rules directly translate into the following Prolog program
that refers to the previously defined {\tt member} relation.

\begin{verbatim}
:- op(1100, yfx, arrow).

% type(E, S, T):-  lambda term S has type T in the environment E.
  type(E, var(X), T):-  member([X, T], E).
  type(E, apply(M, N), T):-  type(E, M, S arrow T), type(E, N, S).
  type(E, lambda(X, M), (S arrow T)):-  type([[X, S] | E], M, T).
\end{verbatim}

For readability we use here \texttt{arrow} as a binary function symbol
written in infix notation.  The first line declares this use of
\texttt{arrow} together with a certain associativity and priority
information (The details of this archaic, though useful, Prolog
notation are not relevant here.)

As expected, the above program can be used to check whether a given
(representation of a) lambda term has a given type.  Less expected is
that this program can also be used to compute a type assignment to a
lambda term, if such an assignment exists, and to report a failure if
no such assignment exists. To this end, given a lambda term $s$, it
suffices to use the query {\tt type([], t, T).}, where the empty list
{\tt []} denotes the empty environment and where {\tt t} is the
translation of $s$ to a first-order term.  For instance, the query
\medskip

\texttt{?- type([], lambda(x, apply(var(x), var(x))), T).}
\medskip

\noindent
fails. In fact, no type can be assigned to the lambda term $\lambda x. \: (x \ x)$. 
The computation first leads to the call
\medskip

\texttt{type([[x, S]], apply(var(x), var(x)), T)} 
\medskip

\noindent
and then to the call
\medskip

\texttt{type([[x, S]], var(x), S arrow T)}.

\medskip

\noindent
This in turn leads to the call
\medskip

\texttt{member([x, S arrow T], [[x, S]])}
\medskip

\noindent
which fails since the terms \texttt{S arrow T} and \texttt{S} do not unify.
In the above computation \texttt{T} is used as a logical variable.

The problem of computing a type assignment for lambda terms was posed
and solved by Curry (see Curry and Feys \cite{CF58}). It is an
important topic in the theory of lambda calculus that is of relevance
for type inference in functional programming.  The solution in Prolog
given above is completely elementary.  A typical use of this program
does not involve backtracking. In fact, its power relies on
unification.

\subsubsection*{A Sequence Program}

Next, consider the following problem:
arrange three 1s, three 2s, ..., three 9s
in sequence so that for all $i \in [1,9]$ there
are exactly $i$ numbers between successive occurrences
of $i$.
An example of such a sequence is
\begin{center}
1, 9, 1, 2, 1, 8, 2, 4, 6, 2, 7, 9, 4, 5, 8, 6, 3, 4, 7, 5, 3, 9, 6, 8, 3, 5, 7.
\end{center}

The desired program is an almost verbatim
formalization of the problem in Prolog.
\begin{verbatim}
% sequence(Xs) :- Xs is a list of 27 variables.
  sequence([_,_,_,_,_,_,_,_,_,_,_,_,_,_,_,_,_,_,_,_,_,_,_,_,_,_,_]).

% question(Ss) :- Ss is a solution to the problem.
  question(Ss) :-
          sequence(Ss),
          sublist([9,_,_,_,_,_,_,_,_,_,9,_,_,_,_,_,_,_,_,_,9], Ss),
          sublist([8,_,_,_,_,_,_,_,_,8,_,_,_,_,_,_,_,_,8], Ss),
          sublist([7,_,_,_,_,_,_,_,7,_,_,_,_,_,_,_,7], Ss),
          sublist([6,_,_,_,_,_,_,6,_,_,_,_,_,_,6], Ss),
          sublist([5,_,_,_,_,_,5,_,_,_,_,_,5], Ss),
          sublist([4,_,_,_,_,4,_,_,_,_,4], Ss),
          sublist([3,_,_,_,3,_,_,_,3], Ss),
          sublist([2,_,_,2,_,_,2], Ss),
          sublist([1,_,1,_,1], Ss).
\end{verbatim}
Note how the anonymous variables dramatically improve the readability of the program.

Operationally, the query \texttt{?- question(Ss).} leads to the
procedure call \texttt{sequence(Ss)} that instantiates the variable
\texttt{Ss} to the list of 27 anonymous (so different) variables.
Then each of the nine calls of the \texttt{sublist} procedure
enforces an existence of a specific sublist pattern on \texttt{Ss}.
Each pattern involves syntactically anonymous variables, each of them
representing operationally a logical variable.

In spite of the fact that the program is simple and transparent 
the resulting computation is involved because of a extensive
use of backtracking. The query generates all six solutions
to the problem.

\subsubsection*{Difference Lists}

One of the drawbacks of the concatenation of lists performed by the
{\tt APPEND} program is that for lists {\tt s} and {\tt t} the execution of the
query {\tt append(s, t, Z)} takes the number of steps that is
proportional to the length of the first list, {\tt s}. This is
obviously inefficient. In an imperative setting if one represents a
list as a link list, to concatenate two lists it suffices to adjust
one pointer.

{\em Difference list\/} is a generalization of the concept of a
list that allows us to perform  concatenation in constant time.
The fact that many programs  rely explicitly on list
concatenation explains the importance of this concept.

In what follows we use the subtraction operator
``{\tt -}'' written in the infix form. Its use has nothing to do 
with arithmetic, though intuitively one should
read it as the ``difference''. 
Formally, a {\em difference list\/} is a construct of the form
${\tt [a_1, ..., a_m | x] - x}$, where {\tt x} is a variable and 
where we used the notation introduced in Subsection \ref{subsec:computing}.
It {\em represents\/} the list ${\tt [a_1, ..., a_m]}$ in a form amenable to a different
definition of concatenation.
Namely, consider two difference lists 
${\tt [a_1, ..., a_m | x] - x}$ and ${\tt [b_1, ..., b_n | y] - y}$.
Then their concatenation is the difference list 
${\tt [a_1, ..., a_m, b_1, ..., b_n | y] - y}$.

This concatenation process is achieved by the following one line {\tt APPEND\_DL} program:

\begin{verbatim}
% append(Xs, Ys, Zs) :- the difference list Zs is the result of 
%                       concatenating the difference lists Xs and Ys.
  append_dl(X-Y, Y-Z, X-Z). 
\end{verbatim}

For example, we have:
\begin{verbatim}
?- append_dl([a,b|X]-X, [c,d|Y]-Y, U). 

U = [a,b,c,d|Y]-Y,
X = [c,d|Y]
\end{verbatim}
which shows that {\tt U} became instantiated to the difference list
representing the list {\tt [a,b,c,d]}.
We shall illustrate the use of difference lists in Subsection \ref{subsec:acr}.

\section{Arithmetics in Prolog}
\label{sec:arithmetic}

The Prolog programs presented so far are \emph{declarative} since
they admit a dual reading as a formula.
The treatment of arithmetic in Prolog compromises to some
extent its declarative underpinnings. However, it is difficult
to come up with a better solution than the one offered by
the original designers of the language.
The shortcomings of Prolog's treatment of arithmetic are overcome
in the constraint logic programming languages.

\subsection{Arithmetic Operators}
\label{subsec:arithmetic-operators}

Prolog provides integers and floating point numbers as built-in data
structures, with the typical operations on them. These operations
include the usual arithmetic operators such as {\tt +}, {\tt -},
{\tt *} (multiplication), and {\tt //} (integer division).

Now, according to the usual notational convention of logic programming and
Prolog, the relation and function symbols are written in the
{\em prefix form\/}, \index{prefix form} that is in front of the arguments. In
contrast, in accordance with their usage in arithmetic, the binary
arithmetic operators are written in {\em infix form\/},
\index{infix form} that is between the arguments.
Moreover,
negation of a natural number can be written in the {\em bracket-less
  prefix form}, \index{bracket-less prefix form} that is without brackets
surrounding its argument.

This discrepancy in the syntax is resolved by considering the
arithmetic operators as built-in function symbols written in the infix
or bracket-less prefix form with information about their associativity
and binding power that allows us to disambiguate the arithmetic
expressions.

Actually, Prolog provides a means to declare an {\em arbitrary\/}
function symbol as an infix binary symbol or as a bracket-less prefix
unary symbol, with a fixed {\em priority\/} \index{priority} that
determines its binding power and a certain {\em mnemonics\/} that
implies some (or no) form of associativity.  An example of such a
declaration was the line \texttt{:- op(1100, yfx, arrow).} used in the
above type assignment program. 
Function symbols declared in this way are called
{\em operators}. \index{operator}
Arithmetic operators can be thus viewed as operators
predeclared in the language ``prelude''. 

In addition to the arithmetic operators we also have at our disposal 
infinitely many integer constants
and infinitely many floating point numbers. 
In what follows by a {\em number} we mean either an integer
constant or a floating point number.
The arithmetic operators and the set of all numbers uniquely
determine a set of terms.  We call terms defined in this language {\em
  arithmetic expressions\/} \index{expression!arithmetic} and
introduce the abbreviation {\em gae\/} \index{gae (ground arithmetic
  expression)} for ground (i.e., variable free) arithmetic
expressions.

\subsection{Arithmetic Comparison Relations}
\label{subsec:acr}

With each gae we can uniquely associate its {\em value},
\index{gae!value of} computed in the expected way.  Prolog allows us
to compare the values of gaes by means of the customary six {\em
  arithmetic comparison relations\/} \index{arithmetic!comparison
  relation}
\medskip

\verb+ <+, 
\verb+ =<+,
\verb+ =:=+ (``equal''),
\verb+ =\=+, (``different''),
\verb+ >=+,
and \verb+ >+.

\medskip

\noindent
The ``equal'' relation ``\verb+ =:=+'' should not be confused with the ``is
unifiable with'' relation ``='' discussed in Section
\ref{sec:unification}.

The arithmetic comparison relations work on gaes and produce the expected outcome.
For instance, \verb+ >+ compares the values of two gaes and succeeds if
the value of the first argument is larger than the value of the second and
fails otherwise.
Thus, for example

\begin{verbatim}
?- 6*2 =:= 3*4. 

yes

?- 7 > 3+4.

no
\end{verbatim}

However, when one of the arguments of the arithmetic comparison relations
is  not a gae, the computation
{\em ends in an error\/}. 
For example, we have

\begin{verbatim}
?- [] < 5.

error in arithmetic expression: [] is not a number.
\end{verbatim}

As a simple example of the use of the arithmetic comparison relations,
consider the following program which checks whether a list of numbers
is ordered.

\begin{verbatim}
% ordered(Xs) :- Xs is an =<-ordered list of numbers
  ordered([]).
  ordered([_]).
  ordered([X, Y | Xs]) :- X =< Y, ordered([Y | Xs]).
\end{verbatim}

Recall that \texttt{[X, Y | Xs])} is the abbreviated Prolog notation for
\texttt{[X | [Y | Xs]])}.
We now have
\begin{verbatim}
?- ordered([1,1,2,3]).

yes
\end{verbatim}
but also
\begin{verbatim}
?- ordered([1,X,1]).

instantiation fault in 1 =< X
\end{verbatim}

Here a run-time error took place because at a certain stage the
comparison relation ``\verb+ =<+'' was applied to an argument that is not
a number. 

As another example consider Prolog's version of the {\em quicksort\/} procedure
of C.A.R. Hoare \index{Hoare, C.A.R.}.
According to this sorting procedure, 
a list is first partitioned into two sublists using an element {\tt X} of it, 
one consisting of the elements smaller than {\tt X} and the other consisting
of the elements larger or equal than {\tt X}.
Then each sublist is quicksorted and the resulting
sorted sublists are appended with the element {\tt X} put in the middle.
This can be expressed in Prolog by means of the following 
{\tt QUICKSORT} program, where {\tt X} is chosen to be the
first element of the given list:

\begin{verbatim}
% qs(Xs, Ys) :- Ys is an ordered permutation of the list Xs.
  qs([], []).
  qs([X | Xs], Ys) :-
          part(X, Xs, Littles, Bigs),
          qs(Littles, Ls),
          qs(Bigs, Bs),
          append(Ls, [X | Bs], Ys).

% part(X, Xs, Ls, Bs) :- Ls is a list of the elements of Xs which are < X, 
%                        Bs is a list of the elements of Xs which are >= X.
  part(_, [], [], []).
  part(X, [Y | Xs], [Y | Ls], Bs) :- X > Y, part(X, Xs, Ls, Bs).
  part(X, [Y | Xs], Ls, [Y | Bs]) :- X =< Y, part(X, Xs, Ls, Bs).
\end{verbatim}

We now have for example
\begin{verbatim}
?- qs([7,9,8,1,5], Ys).

Ys = [1, 5, 7, 8, 9]
\end{verbatim}
and also
\begin{verbatim}
?- qs([7,9,8,1,5], [1,5,7,9,8]).

no
\end{verbatim}

The {\tt QUICKSORT} program uses the {\tt append} relation to
concatenate the lists. Consequently, its efficiency can be improved
using the difference lists introduced in Subsection
\ref{subsec:logical}.  Conceptually, the calls of the {\tt
  append} relation are first replaced by the corresponding calls of the {\tt
  append\_dl} relation.  This yields a program defining the {\tt qs\_dl}
relation.  Then unfolding the calls of {\tt append\_dl}
leads to a program that does not use the {\tt APPEND\_DL} program
anymore and performs the list concatenation ``on the fly''.  
This results in the program {\tt QUICKSORT\_DL} in which the definition of the
{\tt qs} relation is replaced by:

\begin{verbatim}
% qs(Xs, Ys) :- Ys is an ordered permutation of the list Xs.
  qs(Xs, Ys)  :- qs_dl(Xs, Ys - []).
% qs_dl(Xs, Y) :-  Y is a difference list representing the 
%                  ordered permutation of the list Xs.
  qs_dl([], Xs - Xs). 
  qs_dl([X | Xs], Ys - Zs) :-
     part(X, Xs, Littles, Bigs),
     qs_dl(Littles, Ys - [X | Y1s]),
     qs_dl(Bigs, Y1s - Zs). 
\end{verbatim}
The first rule links the {\tt qs} relation with the {\tt qs\_dl} relation.

\subsection{Evaluation of Arithmetic Expressions}
\label{subsec:eval}

So far we have presented programs that use ground arithmetic
expressions, but have not yet introduced any means of evaluating them.
For example, no facilities have been introduced so far to evaluate
{\tt 3+4}. All we can do at this stage is to check that the outcome is
{\tt 7} by using the comparison relation {\tt =:=} and the query {\tt
  7 =:= 3+4}.  But using the comparison relations it is not possible
to {\em assign\/} the value of {\tt 3+4}, that is {\tt 7}, to a
variable, say {\tt X}. Note that the query {\tt X =:= 3+4.} ends in an
error, while the query {\tt X = 3+4.} instantiates {\tt X} to the term
{\tt 3+4}.
 
To overcome this problem
the binary {\em arithmetic evaluator \/} \index{arithmetic!evaluator}
{\tt is} is used in Prolog.
{\tt is} is an infix operator defined as follows.
Consider the call {\tt s is t}. 
Then {\tt t} has to be a ground arithmetic expression (gae).
The call of {\tt s is t} results in the unification of
the {\em value\/} of the gae {\tt t} with {\tt s}.
If {\tt t} is not a gae then a run-time error arises.
So for example we have
\begin{verbatim}
?- 7 is 3+4.   

yes

8 is 3+4.

no 

?- X is 3+4.

X = 7

?- X is Y+1.

! Error in arithmetic expression: not a number
\end{verbatim}

As an example of the use of an arithmetic evaluator
consider the proverbial factorial function. It can be computed using the 
following program {\tt FACTORIAL}:

\begin{verbatim}
% factorial(N, F) :- F is N!.
  factorial(0, 1).
  factorial(N, F) :- N > 0, N1 is N-1, factorial(N1, F1), F is N*F1.
\end{verbatim}

Note the use of a local variable {\tt N1} in the atom
{\tt N1 is N-1} to compute the decrement of {\tt N} and the use of a
local variable {\tt F1} to compute the value of {\tt N1} factorial.
The atom {\tt N1 is N-1} corresponds to the assignment command {\tt N := N-1}
of imperative programming. 
The difference is that a new variable needs
to be used to compute the value of \texttt{N-1}.  Such uses of local
variables are typical when computing with integers in Prolog.

As another example consider a Prolog program
that computes the length of a list.

\begin{verbatim}
% length(Xs, N) :- N is the length of the list Xs.
  length([], 0).
  length([_ | Ts], N) :- length(Ts, M), N is M+1.
\end{verbatim}
We then have
\begin{verbatim}
?- length([a,b,c], N).

N = 3
\end{verbatim}

An intuitive but incorrect version would use as the second
clause
\begin{verbatim}
  length([_ | Ts], N+1) :- length(Ts, N).
\end{verbatim}
With such definition we would get the following
nonintuitive outcome:
\begin{verbatim}
?- length([a,b,c], N).

N = 0 + 1 + 1 + 1
\end{verbatim}
The point is that the generated ground arithmetic expressions
are not automatically evaluated in Prolog.

We conclude that arithmetic facilities in Prolog are quite subtle
and require good insights to be properly used.

\section{Control, Ambivalent Syntax and Meta-Variables}
\label{sec:control}

In the framework discussed so far no control constructs are present.
Let us see now how they could be simulated by means of the 
features explained so far. Consider the customary 
\textbf{if} \texttt{B} \textbf{then} \texttt{S} \textbf{else} \texttt{T} \textbf{fi} 
construct. It can be modelled by means of the following two clauses:
\medskip

{\texttt{p}(\textbf{x}) \texttt{:- B, S}.}

\texttt{p}(\textbf{x}) \texttt{:- not B, T.}

\medskip

\noindent
where \texttt{p} is a new procedure identifier and all the variables
of \texttt{B, S} and \texttt{T} are collected in \textbf{x}.  To see
how inefficiency creeps into this style of programming consider two
cases.

First, suppose that the first clause is selected and that \texttt{B} is
true (i.e., succeeds). Then the computation continues with \texttt{S}.
But in general \texttt{B} is an arbitrary query and because of the
implicit nondeterminism present \texttt{B} can succeed in many ways.
If the computation of \texttt{S} fails, these alternative ways of
computing \texttt{B} will be automatically tried even though we know already
that \texttt{B} is true.

Second, suppose that the first clause is selected and that \texttt{B}
is false (that is fails). Then backtracking takes place and the
second clause is tried.  The computation proceeds by evaluating
\texttt{not B}. This is completely unneeded since we know at this
stage that \texttt{not B} is true (that is succeeds).  

Note that omitting \texttt{not B} in the second rule would cause a
problem in case a success of \texttt{B} were followed by a failure of
\texttt{S}. Then upon backtracking \texttt{T} would be executed.

\subsection{Cut}

To deal with such problems Prolog provides a low level built-in nullary
relation symbol called \emph{cut} and denoted by ``!''.
To explain its meaning we rewrite first the above clauses using cut:
\medskip

{\texttt{p}(\textbf{x}) \texttt{:- B, !, S}.}

{\texttt{p}(\textbf{x}) \texttt{:- T}.}
\medskip

\noindent
In the resulting analysis two possibilities arise, akin to the above
case distinction.  First, if \texttt{B} is true (i.e., succeeds), then
the cut is encountered. Its execution
\begin{itemize}
\item discards all alternative ways of computing \texttt{B},

\item discards the second clause, {\texttt{p}(\textbf{x}) \texttt{:- T}.}, 
as a backtrackable alternative to the current selection of the first clause.
\end{itemize}
Both items have an effect that in the current computation some clauses are not
anymore available.

Second, if \texttt{B} is false (i.e., fails), then backtracking
takes place and the second clause is tried.  The computation proceeds now by
directly evaluating \texttt{T}.
So using the cut and the above rewriting we achieved the intended
effect and modelled the \textbf{if} \texttt{B} \textbf{then}
\texttt{S} \textbf{else} \texttt{T} \textbf{fi} construct in the
desired way.

The above explanation of the effect of cut is a good starting point
to provide its definition in full generality.
Consider the following definition of a relation {\tt p}:
\begin{center}\begin{tabular}{l}
{\tt p($\Seq{s}_1$) :- ${\bf A}_1$}. \\
\dots \\
{\tt p($\Seq{s}_i$)   :- {\bf B},!,{\bf C}}. \\
\dots \\
{\tt p($\Seq{s}_k$) :- ${\bf A}_k$}. \\
\end{tabular}\end{center}
Here, the $i$-th clause contains a cut atom. Now, suppose that during
the execution of a query a call  {\tt p($\Seq{t}$)} is encountered and
eventually the $i$-th clause is used and the indicated occurrence of
the cut is executed.  Then the indicated occurrence
of !  succeeds immediately, but additionally
\begin{enumerate}
\item
all alternative ways of computing $\Seq{B}$ are discarded, and
\item all computations of {\tt p($\Seq{t}$)} using the $i+1$-th to
  $k$-th clause for {\tt p} are discarded as backtrackable
  alternatives to the current selection of the $i$-clause.
\end{enumerate}

The cut was introduced to improve the implicit control present through the
combination of backtracking and the textual ordering of the clauses.
Because of the use of patterns in the clause heads the potential
source of inefficiency can be sometimes hidden somewhat deeper in the
program text.  Reconsider for example the {\tt QUICKSORT} program of
Section \ref{sec:arithmetic} and the query \texttt{?- qs([7,9,8,1,5],
  Ys)}.  To see that the resulting computation is inefficient note
that the second clause defining the \texttt{part} relation fails when 7
is compared with 9 and subsequently the last, third, clause is tried. At
this moment 7 is again compared with 9. The same redundancy occurs
when 1 is compared with 5.  To avoid such inefficiencies the
definition of \texttt{part} can be rewritten using cut as follows:

\begin{verbatim}
  part(_, [], [], []).
  part(X, [Y | Xs], [Y | Ls], Bs) :- X > Y, !, part(X, Xs, Ls, Bs).
  part(X, [Y | Xs], Ls, [Y | Bs]) :- part(X, Xs, Ls, Bs).
\end{verbatim}
Of course, this improvement can be also applied to 
the {\tt QUICKSORT\_DL} program.

Cut clearly compromises the declarative reading of the Prolog
programs.  It has been one of the most criticized features of Prolog.
In fact, a proper use of cut requires a good understanding of Prolog's
computation mechanism and a number of thumb rules were developed to
help a Prolog programmer to use it correctly.  A number of
alternatives to cut were proposed. The most interesting of them,
called \emph{commit}, entered various constraint and parallel logic
programming languages but is not present in standard Prolog.

\subsection{Ambivalent Syntax and Meta-variables}
\label{subsec:meta-variables}

Before we proceed let us review first the basics of Prolog syntax mentioned so far:

\begin{itemize}
\item Variables are denoted by strings starting with an upper case
  letter or ``\_'' (underscore).  In particular, Prolog allows
  so-called anonymous variables, written as ``\_'' (underscore).
  
\item Relation symbols (procedure identifiers), function symbols
  and non-numeric constants are denoted by strings starting with a
  lower case letter.

\item Binary and unary function symbols can be declared 
as infix or bracket-less prefix operators.

\end{itemize}

Now, in contrast to first-order logic, in Prolog the {\em same\/} name
can be used both for function symbols and for relation symbols.
Moreover, the same name can be used for function or relation symbols
of different arity.  This facility is called {\em ambivalent
  syntax\/}. \index{ambivalent syntax} A function or a relation symbol
{\tt f} of arity {\tt n} is then referred to as {\tt f/n}.  So in a
Prolog program we can use both a relation symbol {\tt p/2} and
function symbols {\tt p/1} and {\tt p/2} and build syntactically legal
terms or atoms like {\tt p(p(a,b),c,p(X))}.  

In presence of the ambivalent syntax, the distinction between function
symbols and relation symbols and between terms and
atoms disappears, but in the context of queries and clauses it is
clear which symbol refers to which syntactic category.  

The ambivalent syntax together with Prolog's facility to declare
binary function symbols (and thus also binary relation symbols) as
infix operators allows us to pass queries, clauses and programs as
arguments. In fact, ``{\tt :-}/2'' is declared internally as an infix
operator and so is the comma ``,/2'' between the atoms, so each clause
is actually a term. This facilitates {\em meta-programming}, that is,
writing programs that use other programs as data.

In what follows we shall explain how meta-programming can be realized
in Prolog. To this end we need to introduce one more syntactic feature.
Prolog permits the use of variables in the positions of atoms, both in
the queries and in the clause bodies. Such a variable is called then a
\index{meta-variable} {\em meta-variable\/}.  Computation in the
presence of the meta-variables is defined as before since
the mgus employed can also bind the meta-variables.  So, for
example, given the legal, albeit unusual, Prolog program (that uses
the ambivalent syntax facility)
\begin{verbatim}
  p(a).
  a.
\end{verbatim}
the execution of the Prolog query {\tt p(X), X.} first leads to the
query {\tt a.} and then succeeds.  Here {\tt a} is both a constant and
a nullary relation symbol.

Prolog requires that the meta-variables are properly instantiated before
they are executed. That is, they need to evaluate to a non-numeric term
at the moment they are encountered in an execution. Otherwise a
run-time error arises. For example, for the above program and the
query {\tt p(X), X, Y.} the Prolog computation ends up in error once
the query {\tt Y.} is encountered.

\subsection{Control Facilities}

Let us see now how the ambivalent syntax in conjunction with
meta-variables supports meta-programming.  In this section we limit
ourselves to (meta-)programs that show how to introduce new control
facilities.  We discuss here three examples, each introducing a
control facility actually available in Prolog as a built-in.
More meta-programs will be presented in the next section
once we introduce other features of Prolog.

\paragraph{Disjunction}
To start with we can define disjunction \index{disjunction} by means
of the following simple program:
\begin{verbatim}
  or(X,Y) :- X. 
  or(X,Y) :- Y.
\end{verbatim}
A typical query is then \texttt{or(Q,R)}, where \texttt{Q} and
\texttt{R} are ``conventional queries''.  Disjunction is a Prolog's
built-in declared as an infix operator ``{\tt ;}/2'' and defined by
means of the above two rules, with ``{\tt or}'' replaced by ``;''. So
instead of \texttt{or(Q,R)} one writes \texttt{Q ; R}.

\paragraph{if-then-else}
The other two examples involve the cut operator.
The already discussed \textbf{if} \texttt{B} \textbf{then}
\texttt{S} \textbf{else} \texttt{T} \textbf{fi} construct can be
introduced by means of the by now familiar program
\begin{verbatim}
  if_then_else(B, S, T) :-  B,!,S.
  if_then_else(B, S, T) :-  T.
\end{verbatim}

In Prolog {\tt if\_then\_else} is a built-in defined internally by the
above two rules.  {\tt if\_then\_else(B, S, T)} is written as {\tt B
  -> S;T}, where ``{\tt \ra}/2'' is a
built-in declared as an infix operator.
As an example of its use let us rewrite yet again the definition of the {\tt
  part} relation used in the {\tt QUICKSORT} program, this time using Prolog's
{\tt B -> S;T}.  To enforce the correct parsing we need to enclose the
{\tt B -> S;T} statement in brackets:

\begin{verbatim}
  part(_, [], [], []).
  part(X, [Y | Xs], Ls, Bs) :- 
     ( X > Y ->
       Ls = [Y | L1s], part(X, Xs, L1s, Bs)
     ;
       Bs = [Y | B1s], part(X, Xs, Ls, B1s)
     ).
\end{verbatim}

Note that we had to dispense here with the use of patterns in the
``output'' positions of {\tt part} and reintroduce the explicit use
of unification in the procedure body.  By introducing yet another {\tt
  B -> S;T} statement to deal with the case analysis in the second
argument we obtain a definition of the {\tt part} relation that very
much resembles a functional program:
\begin{verbatim}
  part(X, X1s, Ls, Bs) :- 
     ( X1s = [] ->
       Ls = [], Bs = []
     ;
       X1s = [Y | Xs], 
       ( X > Y ->
         Ls = [Y | L1s], part(X, Xs, L1s, Bs)
       ;
         Bs = [Y | B1s], part(X, Xs, Ls, B1s)
       )
     ).
\end{verbatim}
In fact, in this program all uses of unification boil down to matching and 
its use does not involve backtracking.
This example explains how the use of patterns often hides an implicit
case analysis. By making this case analysis explicit using the {\bf
  if-then-else} construct we end up with longer programs.
In the end the original solution with the cut seems to be closer
to the spirit of the language.
\paragraph{Negation}

Finally, consider the negation operation {\tt not} that is
supposed to reverse failure with success. That is, the intention
is that the query {\tt not Q.} succeeds iff the query {\tt Q.} fails.
This operation can be easily implemented by means of meta-variables and cut
as follows:

\begin{verbatim}
  not(X) :- X, !, fail.
  not(_).
\end{verbatim}
{\tt fail}/0 is Prolog's built-in with the empty definition. Thus
the call of the parameterless procedure {\tt fail} always fails.

This cryptic two line program employs several discussed features of
Prolog.  In the first line {\tt X} is used as a meta-variable. 
Consider now the call {\tt not(Q)}, where {\tt Q} is a query.
If {\tt Q} succeeds, then the cut is performed. This has the effect that all
alternative ways of computing {\tt Q} are discarded and also the
second clause is discarded. Next the built-in {\tt fail} is executed and
a failure arises. Since the only alternative clause was just discarded, the
query {\tt not(Q)} fails.  If on the other hand the query {\tt Q}
fails, then backtracking takes place and the second clause, {\tt
  not(\_)} is selected. It immediately succeeds and so the initial query
{\tt not(Q)} succeeds. So this definition of {\tt not} achieves the desired effect.

{\tt not/1} is defined internally by the above two line
definition augmented with the appropriate declaration of
it as a bracket-less prefix unary symbol.

\paragraph{Call}
Finally, let us mention that Prolog also provides an indirect way of
using meta-variables by means of a built-in relation {\tt call/1}.
{\tt call/1} is defined internally by the rule

\begin{verbatim}
  call(X) :- X.  
\end{verbatim}
{\tt call/1} is often used to ``mask'' the explicit use of meta-variables,
but the outcome is the same.

\subsection{Negation as Failure}

The distinction between successful and failing computations
is one of the unique features of logic programming and Prolog.
In fact, no counterpart of failing computations exists
in other programming paradigms.

The most natural way of using failing computations is by employing the
negation operator \texttt{not} that allows us to turn failure into
success, by virtue of the fact that the query \texttt{not Q.} succeeds
iff the query \texttt{ Q.} fails.  This way we can use \texttt{not} to
represent negation of a Boolean expression. In fact, we already
referred informally to this use of negation at the beginning of
Section \ref{sec:control}.

This suggests a declarative interpretation of the \texttt{not}
operator as a classical negation. This interpretation is correct only
if the negated query always terminates and is ground.  Note in
particular that given the procedure \texttt{p} defined by the single
rule \texttt{p :- p.} the query \texttt{not p.} does not terminate.
Also, for the query \texttt{not(X = 1).} we get the following
counterintuitive outcome:

\begin{verbatim}
?- not(X = 1).

no
\end{verbatim}
Thus to generate all elements of a list \texttt{Ls} that differ from 1
the correct query is \texttt{member(X, Ls), not(X = 1).} and not
\texttt{not(X = 1), member(X, Ls).}

One usually refers to the way negation is used in Prolog as ``negation
as failure''.  When properly used it is a powerful feature as testified
by the following jewel program.  
We consider the problem of determining a winner in a two-person finite
game.  Suppose that the moves in the game are represented by a
relation \texttt{move}. The game is assumed to be finite, so we
postulate that given a position \texttt{pos} the query
\texttt{move(pos, Y).} generates finitely many answers which are all
possible moves from \texttt{pos}.  A player loses if he is in a
position \texttt{pos} from which no move exists, i.e., if the query
\texttt{move(pos, Y).}  fails.

A position is a winning one when a move exists which
leads to a losing, i.e., non-winning position. Using the negation
operator this can be written as
\begin{verbatim}
% win(X) :- X is a winning position in the two-person finite game 
%           represented by the relation move.
  win(X) :- move(X, Y), not win(Y).
\end{verbatim}
So this remarkably concise program has a simple declarative
interpretation.  In contrast, the procedural interpretation is quite
complex: the query \texttt{win(pos).} determines whether \texttt{pos}
is a winning position by performing a minimax search on the 0-1 game
tree represented by the relation \texttt{move}. In this recursive
procedure the base case appears when the call to \texttt{move}
fails---then the corresponding call of \texttt{win} also fails.

\subsection{Higher-Order Programming and Meta-Programming in Prolog}

Thanks to the ambivalent syntax and meta-variables higher-order
programming and another form of meta-programming can be easily
realized in Prolog.  To explain this we need two more built-ins.
Each of them belongs to a different category.

\subsubsection*{Term Inspection Facilities}
\label{subsec:tif}

Prolog offers a number of built-in relations that allow us to inspect,
compare and decompose terms.  One of them is \texttt{=../2} (pronounced
\emph{univ}) that allows us to switch between a term and its representation as
a list.  Instead of describing precisely its meaning we just
illustrate one of its uses by means the following query:

\begin{verbatim}
?- Atom =.. [square, [1,2,3,4], Ys].

Atom = square([1,2,3,4], Ys).
\end{verbatim}
So the left-hand side, here \texttt{Atom}, is unified with the term
(or, equivalently, the atom), here \texttt{square([1,2,3,4], Ys)},
represented by a list on the right-hand side, here \texttt{[square,
  [1,2,3,4], Ys]}. In this list representation of a term the head of
the list is the leading function symbol and the tail is the list of the
arguments.

So using \emph{univ} one can construct terms and pass them as arguments.
More interestingly, one can construct atoms and execute them using the
meta-variable facility. This way it is possible to realize
higher-order programming in Prolog in the sense that relations can be
passed as arguments.  To illustrate this point consider the following
program \texttt{MAP}:

\begin{verbatim}
% map(P, Xs, Ys) :- the list Ys is the result of applying P 
%                   elementwise to the list Xs.
  map(P, [], []).
  map(P, [X | Xs] , [Y | Ys]) :- apply(P, [X, Y]), map(P, Xs, Ys).

% apply(P, [X1, ..., Xn]) :- execute the atom P(X1, ..., Xn).
  apply(P, Xs) :- Atom =.. [P|Xs], Atom.
\end{verbatim}
In the last rule \emph{univ} is used to construct an atom.
Note the use of the meta-variable \texttt{Atom}.
\texttt{MAP} is Prolog's counterpart of the familiar higher-order
functional program and it behaves in the expected way.  For example,
given the program
\begin{verbatim}
% square(X, Y) :- Y is the square of X.
  square(X, Y) :- Y is X*X.
\end{verbatim}
we get

\begin{verbatim}
?- map(square, [1,2,3,4], Ys).

Ys = [1, 4, 9, 16]
\end{verbatim}

\subsubsection*{Program Manipulation Facilities}

Another class of Prolog built-ins makes it possible to access and
modify the program during its execution.  We consider here a single
built-in in this category, {\tt clause/2} \index{{\tt clause/2}}, that
allows us to access the definitions of the relations present in the
considered program. Again, consider first an example of its use in
which we refer to the program {\tt MEMBER} of Subsection
\ref{subsec:multiple}.

\begin{verbatim}
?- clause(member(X,Y), Z).

Y = [X|_A],
Z = true ;

Y = [_A|_B],
Z = member(X,_B) ;

no
\end{verbatim}

In general, the call {\tt clause(head, body)} leads to a unification
of the term {\tt head :- body} with the successive clauses forming the
definition of the relation in question.  This relation, here
\texttt{member}, is the leading symbol of the first argument of
{\tt clause/2} that has to be a non-variable.

This built-in assumes that {\tt true} is the body of a fact, here
\texttt{member(X, [X | \_])}.  {\tt true}/0 is Prolog's built-in that
succeeds immediately. So its definition consists just of the fact
\texttt{true}.  This explains the first answer.  The second answer is
the result of unifying the term \texttt{member(X,Y) :- Z} with (a
renaming of) the second clause defining \texttt{member}, namely
\texttt{member(X, [\_ | Xs]):- member(X, Xs)}.

Using {\tt clause/2} we can construct Prolog interpreters written in
Prolog, that is, \emph{meta-interpreters}.  Here is the simplest one.

\begin{verbatim}
% solve(Q) :- the query Q succeeds for the program accessible by clause/2.
  solve(true) :- !. 
  solve((A,B)) :- !, solve(A), solve(B). 
  solve(A) :- clause(A, B), solve(B).
\end{verbatim}

Recall that {\tt (A,B)} is a legal Prolog term (with no leading function symbol).
To understand this program one needs to know that the comma ``,''
between the atoms is declared internally as a right associative infix
operator, so the query {\tt A,B,C,D} actually stands for the term {\tt
  (A,(B,(C,D)))}, etc.

The first clause states that the built-in {\tt true} succeeds immediately.
The second clause states that a query of the form
$A, \vect{B}$ can be solved if $A$ can be solved and $\vect{B}$
can be solved. Finally, the last clause states that an atomic query
$A$ can be solved if there exists a clause of the form $A :- \ \vect{B}$
such that the query $\vect{B}$ can be solved.
The cuts are used here to enforce the a ``definition by cases'':
either the argument of {\tt solve} is {\tt true} or a non-atomic
query or else an atomic one.

To illustrate the behavior of the above meta-interpreter
assume that {\tt MEMBER} is a part of the considered
program. We then have
\begin{verbatim}
?- solve(member(X, [mon, wed, fri])).

X = mon ;

X = wed ;

X = fri ;

no
\end{verbatim}
This meta-program forms a basis for building various types of interpreters
for larger fragments of Prolog or for its extensions.

\section{Assessment of Prolog}

Prolog, due to its origin in automated theorem proving, is an unusual
programming language.  It leads to a different style of programming
and to a different view of programming.  A number of elegant Prolog
programs presented here speak for themselves. We also noted that the
same Prolog program can often be used for different purposes ---for
computing, testing or completing a solution, or for computing all
solutions.  Such programs cannot be easily written in other
programming paradigms. Logical variables are a unique and, as we saw,
very useful feature of logic programming.  Additionally, pure Prolog
programs have a dual interpretation as logical formulas. In this sense
Prolog supports declarative programming.

Both through the development of a programming methodology and ingenious
implementations a great care was taken to overcome possible sources
of inefficiency.  On the programming level we already discussed cut
and the difference lists.  Programs such as {\tt FACTORIAL} of Subsection
\ref{subsec:eval} can be optimized by means of tail recursion.  On
the implementation level the efficiency is improved by such techniques
as the last call optimization that can be used to optimize tail recursion,
indexing that deals with the presence of multiple clauses, and a default
omission of the occur-check (the test ``$x$ does not occur in $t$'' in
clause (5) of the Martelli--Montanari algorithm) that speeds up the
unification process (though on rare occasions makes it unsound).

Prolog's only data type, the terms, is implicitly present in many
areas of computer science. In fact, whenever the objects of interest
are defined by means of grammars, for example first-order formulas,
digital circuits, programs in any programming language, or sentences
in some formal language, these objects can be naturally defined as
terms.  Prolog programs can then be developed starting with this
representation of the objects as terms.  Prolog's support for handling
terms by means of unification and various term inspection facilities
comes then handy. In short, symbolic data can be naturally handled in
Prolog.

Automatic backtracking becomes very useful when dealing with
search. Search is of paramount importance in many artificial
intelligence applications and backtracking itself is most natural when
dealing with NP-complete problems.  Moreover, the principle of
``computation as deduction'' underlying Prolog's computation process
facilitates formalization of various forms of reasoning in Prolog. In
particular, Prolog's negation operator \texttt{not} can be naturally
used to support non-monotonic reasoning.  All this explains why Prolog
is a natural language for programming artificial intelligence
applications, such as automated theorem provers, expert systems and
machine learning programs where reasoning needs to be combined with
computing, game playing programs, and various decision support
systems.

Prolog is also an attractive language for computational linguistics
applications and for compiler writing. In fact, Prolog provides
support for so-called definite clause grammars (DCG).  Thanks to this
a grammar written in the DCG form is already a Prolog program that
forms a parser for this grammar.  The fact that Prolog allows one to
write executable specifications makes it also a useful language
for rapid prototyping, in particular in the area of meta-programming.

For the sake of a balanced presentation let us discuss now Prolog's shortcomings.

\paragraph{Lack of Types}

Types are used in programming languages to structure the data manipulated
by the program and to ensure its correct use.
In Prolog one can define various types like binary trees and
records.  Moreover, the language provides a notation for lists and offers a
limited support for the type of all numbers by means of the arithmetic
operators and arithmetic comparison relations.  However, Prolog does
not support types in the sense that it does not check whether the
queries use the program in the intended way.  

Because of this absence of type checking, type errors are
easy to make but difficult to find.  For example, even though the
\texttt{APPEND} program was meant to be used to concatenate two lists
it can also be used with non-lists as arguments:
\begin{verbatim}
?-  append([a,b], f(c), Zs). 

Zs = [a, b|f(c)]
\end{verbatim}
and no error is reported.  In fact, almost every Prolog program can be
misused.  Moreover, because of lack of type checking some improvements
of the efficiency of the implementation cannot be carried out and various
run-time errors cannot be prevented.

\paragraph{Subtle Arithmetic}
We discussed already the subtleties arising in presence of arithmetic
in Section \ref{sec:arithmetic}.  We noted that Prolog's facilities
for arithmetic easily lead to run-time errors. It would be desirable
to discover such errors at compile time but this is highly non-trivial.

\paragraph{Idiosyncratic control}

Prolog's control mechanisms are difficult to master by programmers
accustomed to the imperative programming style.  One of the reasons is
that both bounded iteration (the {\tt for} statement) and unbounded
iteration (the {\tt while} statement) need to be implemented by means
of recursion. So for example a nested {\tt for} statement is
implemented by means of nested tail recursion that is less easy to
understand. Of course, one can introduce both constructs by means of
meta-programming but then their proper use is not enforced due to
the lack of types.
Additionally, as already mentioned, cut is a low level mechanism
that is not easy to understand.

\paragraph{Complex semantics of various built-ins}

Prolog offers a large number of built-ins. In fact, the ISO Prolog
Standard \cite{Iso95} describes 102 built-ins.  Several of them are
quite subtle.  For example, the query \texttt{not(not Q).}  tests
whether the query \texttt{Q.} succeeds and this test is carried out
without changing the state, i.e., without binding any of the
variables. Moreover, it is not easy to describe precisely the
meaning of some of the built-ins. For example, in the ISO Prolog
Standard the operational interpretation of the {\bf if-then-else}
construct consists of 17 steps.

\paragraph{No Modules and no Objects}

Finally, even though modules exist in many widely used Prolog
versions, neither modules nor objects are present in ISO Prolog
Standard. This makes it difficult to properly structure Prolog
programs and to reuse them as components of other Prolog programs.  It
should be noted that thanks to Prolog's support for meta-programming
the object-programming style can be mimicked in Prolog in a pretty
simple way. But no compile-time checking of its proper use is then
enforced and errors in the program design will be discovered at
best at the run-time.  The same critique applies to Prolog's approach
to higher-order programming and to meta-programming.  \medskip

Of course, these limitations of Prolog were recognized by many
researchers who came up with various good proposals how to improve
Prolog's control, how to add to it (or how to infer) types, and how to
provide modules and objects. Research in the field of logic
programming has also dealt with the precise relation between the
procedural and declarative interpretation of logic programs and a
declarative account of various aspects of Prolog, including negation
and meta-programming. Also verification of Prolog programs and its
semantics were extensively studied.

However, no single programming language proposal emerged yet that
could be seen as a natural successor to Prolog in which the above
shortcomings are properly taken care of.  The language that comes
closest to this ideal is Mercury (see
\verb+http://www.cs.mu.oz.au/research/mercury/+).  Colmerauer
\index{Colmerauer, A.} himself designed a series of successors of
Prolog, Prolog II, III and IV that incorporated various forms of
constraint processing into this programming style.

When assessing Prolog it is useful to have in mind that it is a
programming language designed in the early seventies (and standardized in
the nineties).  The fact that it is still widely used and that new
applications for it keep being found testifies to its originality.  No
other programming language succeeded to embrace first-order logic in
such an effective way.

\section{Bibliographic Remarks}

For those interested to learn in detail the origins of logic
programming and of Prolog there is no better place to start than to
read the fascinating account in \cite{CR96}.
There a number of excellent books on programming in Prolog.  The two
deservedly most successful are \index{Bratko, I.} Bratko \cite{Bra01}
and \index{Sterling, L.} \index{Shapiro, E.}  Sterling and Shapiro
\cite{SS94}.  The book of O'Keefe \cite{O'K90} \index{O'Keefe, R.A.}
discusses in depth the efficiency and pragmatics of programming in
Prolog.  A{\"{\i}t}-Kaci \cite{Ait91} \index{A{\"{\i}t}-Kaci, H.} is
an outstanding tutorial on the implementation of Prolog.

\section{Summary}
We discussed here the logic programming paradigm and its realization
in Prolog.  This paradigm has contributed a number of novel ideas
in the area of programming languages. It introduced unification as a
computation mechanism and it realized the concept of ``computation as
deduction''.  Additionally, it showed that a fragment of first-order
logic can be used as a programming language and that declarative
programming is an interesting alternative to structured programming in
the imperative programming style.

Prolog is a rule based language but thanks to a large number of
built-ins it is a general purpose programming language.  Programming
in Prolog substantially differs from programming in the imperative
programming style. The following table may help to relate the underlying
concepts used in both programming styles.

\medskip

\begin{center}
  
\begin{tabular}{|l|l|}
\hline
logic programming                  & imperative programming \\ \hline
equation solved by unification    & assignment \\
relation symbol                    & procedure identifier \\
term                               & expression \\
atom                               & procedure call \\
query                              & program \\
definition of a relation           & procedure declaration \\
local variable of a rule           & local variable of a procedure \\
logic program                      & set of procedure declarations \\
``,'' between atoms                & sequencing (``;'')  \\ 
substitution                       & state \\ 
composition of substitutions       & state update \\
\hline
\end{tabular}  
\end{center}
\medskip

\subsection*{Acknowledgements}
Maarten van Emden and Jan Smaus provided us with useful comments on
this article.


\end{document}